\documentclass[aps,prl,preprint,superscriptaddress]{revtex4-1}
\usepackage{graphicx}
\usepackage{dcolumn}
\usepackage{bm}
\usepackage[latin1]{inputenc}
\begin{document}


\title{Radiation-reaction electromagnetic fields in metasurfaces, a complete description of their optical properties}

\author{Michele Merano}
\email[]{michele.merano@unipd.it}
\affiliation{Dipartimento di Fisica e Astronomia G. Galilei, Universit$\grave{a}$ degli studi di Padova, via Marzolo 8, 35131 Padova, Italy}


\date{\today}

\begin{abstract}
This paper derives the macroscopic electric and magnetic fields and the surface susceptibilities for a metasurface, starting from the microscopic scatterer distribution. It is assumed that these scatterers behave as electric and magnetic dipoles under the influence of the incident radiation. Interestingly not only the retarded electromagnetic fields from oscillating dipoles are relevant to pass from the microscopic to the macroscopic representation, but the advanced fields must be considered too. It is found that the macroscopic fields are the sum of the incident fields plus the radiation-reaction fields acting on a single scatterer. Both the local fields and the radiation-reaction fields are necessary to fix the electric and magnetic surface susceptibilities. 
\end{abstract}

\maketitle

\section{INTRODUCTION}
\label{sec:intro}  

A metasurface is a two-dimensional (2D) metamaterial, a material engineered to have a property that is not found in nature. As such, metasurfaces are extremely advanced, manmade, new tools. They consist of a certain density of distributed microscopic scatterers that induce unusual macroscopic reflection and transmission properties \cite{Capasso14, Arbabi15, Shalaev13, Brongersma14}. These scatterers are much smaller than the wavelength of the incident electromagnetic field. They are designed to have well-defined electric and magnetic polarizabilities. 

The connection between the microscopic polarizability and the macroscopic reflected and transmitted fields is not trivial. In the atomic theory for standard three dimensional (3D) materials, the procedure to pass from micro to macro runs as follow \cite{Wolf}. The contribution of the magnetic polarizability is in general neglected because it is small, and the magnetic susceptibility is set equal to zero. To estimate the difference in between the local electric field and the macroscopic electric field, we can imagine that a scatterer (an atom or a molecule) is surrounded by a sphere with a radius big compared with the interatomic distance and we can consider separately the effects produced by the matter outside and inside the sphere. It is shown that the scatterers inside the sphere do not produce any effect on the central one, hence we may regard it as being situated in a spherical region inside which there is vacuum and outside which there is a uniform polarization. This results in the Clausius-Mossotti-Lorenz-Lorentz relation that connects the microscopic atomic electrical polarizability to the macroscopic electrical susceptibility.

Up to now the relation between the microscopic and the macroscopic electromagnetic fields in a metasurface has been done in a manner analogous to this 3D procedure \cite{Kuester03}. Of course in the case of a metasurface the magnetic susceptibility plays an important role, but the magnetic polarizability can be treated in a similar way to the electric one. The important difference going from the bulk to a surface is the replacement of the sphere with a disc. If for the 3D case, it is enough to choose the radius of the sphere big enough, in the 2D case the radius of the disk must be determined in some way. This leads to some arbitrariness in this choice. 

In parallel with the advent of metasurfaces, the last decade has also witnessed the advent of 2D crystals \cite{Novoselov2005, Heinz2010}. These single-layer atomic planes are stable under ambient conditions and appear continuous on a macroscopic scale. Macroscopic continuity is confirmed also by their optical properties that can be essentially described by a surface electric susceptibility \cite{Merano16, Jayaswal18}. Differently to a bulk material and in analogy with a metasurface, in a 2D crystal all the dipoles contribute to the local electric field, and due to the finite velocity of propagation of the dipolar potential, retardation effects manifest themselves up to the macroscopic description \cite{Luca16}. 

Starting from the expression of the local electric field, a recent paper has connected this microscopic description to the macroscopic one by deducing an expression for the macroscopic field in the 2D crystal \cite{Merano17}. Remarkably an essential role in this context is played by the radiation-reaction field acting on a single dipole, via the boundary conditions for the system. The expression of the radiation-reaction electric field coincides with the macroscopic electric field radiating from the crystal and, summed to the incident electric field, generates the total macroscopic electric field. This theory does not require the calculation of the fields of any small disk of arbitrary radius, it is based on a well-defined and relativistically invariant prescription and it is well supported by experiments \cite{Merano16}.

In ref. \cite{Merano17} only the case for a crystal with a magnetic surface susceptibility equal to zero was addressed. Here I show how to extend the same concept to a metasurface, with scatterers that have both electric and magnetic polarizabilities.

\section{CLASSICAL THEORY OF A RADIATING METASURFACE}
\begin{figure} [ht]
   \begin{center}
   \begin{tabular}{c} 
   \includegraphics[height=5cm]{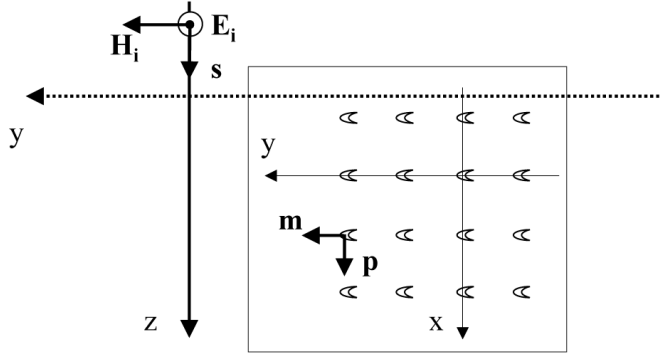}
   \end{tabular}
   \end{center}
   \caption[example] 
   { \label{geometry} 
An $s$ polarized electromagnetic wave is incident on a metasurface, located in the plane $z$ =0. Inset: scatterers distributed on a square two-dimensional Bravais lattice. The scatterers have non-null electric and magnetic polarizabilities. For simplicity I assume that each scatterer behaves as an isotropic oscillating electric and magnetic dipole even if in the more general case metasurfaces are made of anisotropic scatterers.}
   \end{figure}

For simplicity I model a metamaterial as scatterers on a Bravais lattice (fig. (\ref{geometry})). I consider two types of 2D  Bravais lattices, the square and the triangular one. The unit cell surface density ($N$) is  
\begin{equation}
\label{N}
N=1/a^2\quad \textrm{(square lattice)}; \quad  N=2/\sqrt{3}a^2 \quad  \textrm{(triangular lattice)}
\end{equation}
where $a$ is the lattice spacing. An $s$ polarized plane wave is normally incident on a free-standing metasurface. It generates a reflected and a transmitted wave, a macroscopic electric $\vec{\textbf{\emph{E}}}(t)$ and magnetic $\vec{\textbf{\emph{H}}}(t)$ field, a surface polarization $\vec{\textbf{\emph{P}}}(t)$ and a surface magnetization $\vec{\textbf{\emph{M}}}(t)$. These quantities are connected by the following relations
\begin{equation}
\label{E, H}
\vec{\textbf{\emph{P}}}(t)=\epsilon_0 \chi_e \vec{\textbf{\emph{E}}}(t); \quad \vec{\textbf{\emph{M}}}(t)= \chi_m \vec{\textbf{\emph{H}}}(t)  
\end{equation}
where $\epsilon_0$ is the vacuum permittivity, $\chi_e$ and $\chi_m$ are the macroscopic surface electric and magnetic susceptibilities. $\vec{\textbf{\emph{P}}}$ and $\vec{\textbf{\emph{M}}}$ are also related to the microscopic electric ($\alpha_e$) and magnetic ($\alpha_m$) polarizabilities via
\begin{equation}
\label{P, M}
\vec{\textbf{\emph{P}}}(t)=N\vec{\textbf{\emph{p}}}(t)=N\epsilon_0 \alpha_e \vec{\textbf{\emph{E}}}_{loc}(t); \quad \vec{\textbf{\emph{M}}}(t)=N\vec{\textbf{\emph{m}}}(t)=N \alpha_m \vec{\textbf{\emph{H}}}_{loc}(t).
\end{equation}

$\vec{\textbf{\emph{E}}}_{loc}(t)$ and $\vec{\textbf{\emph{H}}}_{loc}(t)$  are the local electric and magnetic fields acting on a single scatterer of the unit cell, $\vec{\textbf{\emph{p}}}(t)$ and $\vec{\textbf{\emph{m}}}(t)$ are the electric and magnetic dipole moments established for each scatterer under the influence of the local fields.

Given the metasurface structure with $N$, $a$, $\alpha_e$, $\alpha_m$ and the incident electric and magnetic fields $\vec{\textbf{\emph{E}}}_i(t)$, $\vec{\textbf{\emph{H}}}_i(t)$, a complete theory must be able to compute the microscopic fields $\vec{\textbf{\emph{E}}}_{loc}(t)$ and $\vec{\textbf{\emph{H}}}_{loc}(t)$, and then the macroscopic fields  $\vec{\textbf{\emph{P}}}(t)$, $\vec{\textbf{\emph{M}}}(t)$, $\vec{\textbf{\emph{E}}}(t)$, $\vec{\textbf{\emph{H}}}(t)$ and $\chi_e$, $\chi_m$. 

\section{Microscopic theory}
The scatterers, excited by the incident fields, act as electric and magnetic dipoles. For simplicity I consider isotropic scatterers, even if in the more general case, metasurfaces are created by assembling arrays of miniature, anisotropic radiation scatterers. If I assume an harmonic time dependence $e^{i(\omega t-k z)}$ for the incident field, an electric dipole generates an electromagnetic field given by \cite{Jackson}
\begin{eqnarray}
\label{electric_dipole_field}
\vec{\textbf{\emph{E}}}_p(\vec{\textbf{\emph{r}}}, t)&=&\frac{e^{i(\omega t-k r)}}{4\pi \epsilon_{0}r^3} \biggl([3(\vec{\textbf{\emph{p}}}_0\cdot\hat{\textbf{\emph{n}}})\hat{\textbf{\emph{n}}}-\vec{\textbf{\emph{p}}}_0]\big(1+ikr \big) +k^2 r^2( \hat{\textbf{\emph{n}}} \times \vec{\textbf{\emph{p}}}_0)  \times \hat{\textbf{\emph{n}}} \biggr) \nonumber \\
\vec{\textbf{\emph{H}}}_p((\vec{\textbf{\emph{r}}}, t)&=&\frac{c k^2 e^{i(\omega t-k r)}}{4\pi r}( \hat{\textbf{\emph{n}}} \times \vec{\textbf{\emph{p}}}_0)\biggl(1+\frac{1}{ikr}\biggl)
\end{eqnarray}
where $c$ is the speed of light, $\hat{\textbf{\emph{n}}}$ is the unit vector in the direction of $\vec{\textbf{\emph{r}}}$ and $\vec{\textbf{\emph{p}}}(t)=\vec{\textbf{\emph{p}}}_0 e^{i\omega t}$. I note that the magnetic field is transverse to the radius vector at all distances. The electromagnetic field generated by the magnetic dipole $\vec{\textbf{\emph{m}}}(t)=\vec{\textbf{\emph{m}}}_0 e^{i\omega t}$ is 
\begin{eqnarray}
\label{magnetic_dipole_field}
\vec{\textbf{\emph{H}}}_m(\vec{\textbf{\emph{r}}}, t)&=&\frac{e^{i(\omega t-k r)}}{4\pi r^3} \biggl([3(\vec{\textbf{\emph{m}}}_0\cdot\hat{\textbf{\emph{n}}})\hat{\textbf{\emph{n}}}-\vec{\textbf{\emph{m}}}_0]\big(1+ikr \big) +k^2 r^2( \hat{\textbf{\emph{n}}} \times \vec{\textbf{\emph{m}}}_0)  \times \hat{\textbf{\emph{n}}} \biggr) \nonumber \\
\vec{\textbf{\emph{E}}}_m((\vec{\textbf{\emph{r}}}, t)&=&-\frac{\eta k^2 e^{i(\omega t-k r)}}{4\pi r}( \hat{\textbf{\emph{n}}} \times \vec{\textbf{\emph{m}}}_0)\biggl(1+\frac{1}{ikr}\biggl)
\end{eqnarray}
and it can be obtained from expressions (\ref{electric_dipole_field}) with the interchange $\vec{\textbf{\emph{E}}}_p \rightarrow \eta \vec{\textbf{\emph{H}}}_m$, $\vec{\eta \textbf{\emph{H}}}_p \rightarrow  -\vec{\textbf{\emph{E}}}_m$ and $\vec{\textbf{\emph{p}}}_0 \rightarrow \vec{\textbf{\emph{m}}}_0/c$ \cite{Jackson}.

\subsection{The local fields}
The local field acting on a scatterer is provided by the superposition principle, being the sum of the incident field plus the electromagnetic fields due to all the other scatterers except the one under consideration.
\begin{eqnarray}
\label{Loc}
\vec{\textbf{\emph{E}}}_{loc}(t)&=&\vec{\textbf{\emph{E}}}_{i}(t)+{\sum_{(n, s)}}'\vec{\textbf{\emph{E}}}_{p \ (n,s)}+{\sum_{(n,s)}}'\vec{\textbf{\emph{E}}}_{m\ (n,s)} \nonumber \\
\vec{\textbf{\emph{H}}}_{loc}(t)&=&\vec{\textbf{\emph{H}}}_{i}(t) +{\sum_{(n,s)}}'\vec{\textbf{\emph{H}}}_{m\ (n,s)}+{\sum_{(n,s)}}'\vec{\textbf{\emph{H}}}_{p\ (n,s)}
\end{eqnarray}
where the sites of the Bravais lattice are labeled by the integers $(n, s)$ and the prime indicates that the scatterer, for which the local field is computed, does not contribute. From the expressions (\ref{electric_dipole_field}, \ref{magnetic_dipole_field}) for simmetry reasons
\begin{equation}
\label{sum_zero}
{\sum_{(n,s)}}'\vec{\textbf{\emph{E}}}_{m\ (n,s)}=0  \  \textrm{V/m}; \quad {\sum_{(n,s)}}'\vec{\textbf{\emph{H}}}_{p\ (n,s)}=0\  \textrm{A/m}.
\end{equation}
Sums (\ref{Loc}) reduce in this way to those already computed in reference \cite{Luca16} and give the following results
\begin{eqnarray}
\label{LocalE}
\nonumber\vec{\textbf{\emph{E}}}_{i}(t)&=&\vec{\textbf{\emph{E}}}_{loc}(t)
\left(1-\frac{\alpha_e C_{0}}{4\pi a^3}+i \frac{\alpha_e N k}{2}
\right)\\
\vec{\textbf{\emph{H}}}_{i}(t)&=&\vec{\textbf{\emph{H}}}_{loc}(t)
\left(1-\frac{\alpha_m C_{0}}{4\pi a^3}+i \frac{\alpha_m N k}{2}
\right)
\end{eqnarray}
where the values of $C_{0}$ for the square and the triangular lattice are reported in \cite{Luca16}.

\section{From Micro to Macro}
The microscopic theory provides $\vec{\textbf{\emph{E}}}_{loc}(t)$ and $\vec{\textbf{\emph{H}}}_{loc}(t)$. From the local fields it is then possible (eqs. \ref{P, M})  to compute $\vec{\textbf{\emph{P}}}(t)$ and $\vec{\textbf{\emph{M}}}(t)$. We still do not know (eqs. \ref{E, H}) both $\vec{\textbf{\emph{E}}}(t)$, $\chi_e$ and $\vec{\textbf{\emph{H}}}(t)$, $\chi_m$. It is clear that something is missing in the theory. For a 2D crystal, reference \cite{Merano17} has shown that the missing term is the radiation-reaction electric field $\vec{\textbf{\emph{E}}}_{R}(t)$ acting on a single scatter. I prove below that the same is true for a metasurface.

\subsection{Radiation-reaction electromagnetic fields}
$\vec{\textbf{\emph{E}}}_{R}(t)$ is the difference between one-half the retarded and one-half the advanced electric fields generated by all the other scatterers except the one under consideration\cite{Merano17, Wheeler45, Dirac38, Panofsky}. In an analogous way this definition is extended to the radiation-reaction magnetic field
\begin{eqnarray}
\label{Reaction_}
\vec{\textbf{\emph{E}}}_{R}(t) &= &{\sum_{(n,s)}}' \frac{1}{2}\bigg( \vec{\textbf{\emph{E}}}_{ret, \ (n,s)}(t)-\vec{\textbf{\emph{E}}}_{adv, \ (n,s)}(t) \bigg) \qquad \nonumber \\ 
\vec{\textbf{\emph{H}}}_{R}(t) &= &{\sum_{(n,s)}}' \frac{1}{2}\bigg( \vec{\textbf{\emph{H}}}_{ret, \ (n,s)}(t)-\vec{\textbf{\emph{H}}}_{adv, \ (n,s)}(t) \bigg).
\end{eqnarray}   
Equations (\ref{electric_dipole_field}, \ref{magnetic_dipole_field}) give the retarded electromagnetic fields generated by oscillating electric and magnetic dipoles. The advanced fields are easily related to the retarded ones \cite{Merano17}, if the retarded fields are given by 
\begin{equation}
\label{retarded}
\vec{\textbf{\emph{E}}}_{ret}(\vec{\textbf{\emph{r}}}, t)= e^{i \omega t}\vec{\textbf{\emph{f}}}(\vec{\textbf{\emph{r}}}); \quad \vec{\textbf{\emph{H}}}_{ret}(\vec{\textbf{\emph{r}}}, t)= e^{i \omega t}\vec{\textbf{\emph{g}}}(\vec{\textbf{\emph{r}}})
\end{equation}
the advanced fields are
\begin{equation}
\label{advanced}
\vec{\textbf{\emph{E}}}_{adv}(\vec{\textbf{\emph{r}}}, t)= e^{i \omega t}\vec{\textbf{\emph{f}}}^*(\vec{\textbf{\emph{r}}}); \quad \vec{\textbf{\emph{H}}}_{adv}(\vec{\textbf{\emph{r}}}, t)= e^{i \omega t}\vec{\textbf{\emph{g}}}^*(\vec{\textbf{\emph{r}}}).
\end{equation}
Keeping in mind eqs. (\ref{sum_zero}, \ref{LocalE}) I obtain
\begin{eqnarray}
\label{Reaction}
\vec{\textbf{\emph{E}}}_{R}(t) &= &-\frac{i\, \alpha_e N k}{2} \vec{\textbf{\emph{E}}}_{loc}(t)=- \frac{\eta}{2}\dot{\vec{\textbf{\emph{P}}}}(t)=-\frac{\eta}{2}\vec{\textbf{J}}_p(t) \nonumber \\ 
\vec{\textbf{\emph{H}}}_{R}(t) &= &-\frac{i\, \alpha_m N k}{2} \vec{\textbf{\emph{H}}}_{loc}(t)= -\frac{\dot{\vec{\textbf{\emph{M}}}}(t)}{2c}=-\frac{1}{2\eta}\vec{\textbf{J}}_m(t)  
\end{eqnarray} 
where $\vec{\textbf{J}}_p=\dot{\vec{\textbf{\emph{P}}}}$ is the surface polarization current and $\vec{\textbf{J}}_m = \mu_0 \dot{\vec{\textbf{\emph{M}}}}$ is the surface magnetization current. From these expressions it is evident that $\vec{\textbf{\emph{E}}}_{R}(t)$ and $\vec{\textbf{\emph{H}}}_{R}(t) $ are macroscopic quantities.

\subsection{Boundary conditions for a metasurface}
Next we consider the amplitudes of the reflected $\vec{\emph{\textbf{E}}}_r$, $\vec{\emph{\textbf{H}}}_r$ and transmitted $\vec{\emph{\textbf{E}}}_t$, $\vec{\emph{\textbf{H}}}_t$ fields. These relations hold
\begin{equation}
\label{EsH}
\eta\vec{\emph{\textbf{H}}}_i=\hat{s}_i\wedge\vec{\emph{\textbf{E}}}_i; \quad \eta\vec{\emph{\textbf{H}}}_r=\hat{s}_r\wedge\vec{\emph{\textbf{E}}}_r; \quad \eta\vec{\emph{\textbf{H}}}_t=\hat{s}_t\wedge\vec{\emph{\textbf{E}}}_t;
\end{equation}
where $\hat{s}_{(i, r, t)}$ are the unit vectors along the respective propagation directions. The boundary conditions demand that across the metasurface the tangential components of the electric and the magnetic fields are discontinuous \cite{Kuester03}
\begin{equation}
\label{Boundary}
\hat{\kappa} \wedge \bigg(\vec{\textbf{E}}\mid_{z+}-\vec{\textbf{E}}\mid_{z-}\bigg)=-\vec{\textbf{J}}_m=2\eta\vec{\textbf{H}}_R; \quad \hat{\kappa} \wedge \bigg(\vec{\textbf{H}}\mid_{z+}-\vec{\textbf{H}}\mid_{z-}\bigg)=\vec{\textbf{J}}_p=-\frac{2}{\eta}\vec{\textbf{E}}_R
\end{equation}
where $\hat{\kappa}$ is the unit vector along the $z$ direction. From eqs. (\ref{Boundary}) I obtain
\begin{equation}
\label{Boundary2}
E_i+E_r=E_t-2\eta H_R; \quad H_i-H_r=H_t-\frac{2}{\eta}E_R.
\end{equation} 
From eqs. (\ref{EsH}) and (\ref{Boundary2}) the amplitude of the reflected and the transmitted fields are
\begin{equation}
\label{Fresnel_micro}
E_r=E_R-\eta H_R; \quad E_t=E_i+E_R+\eta H_R; \quad H_r=\frac{E_R}{\eta}- H_R  ;\quad H_t=H_i+\frac{E_R}{\eta}+ H_R.
\end{equation} 

\subsection{The macroscopic fields}
In analogy with reference \cite{Kuester03}, I define the macroscopic fields as the average fields at the metasurface
\begin{equation}
\label{Boundary}
\vec{\textbf{\emph{E}}}=\frac{1}{2}\bigg(\vec{\textbf{E}}\mid_{z+}+\vec{\textbf{E}}\mid_{z-}\bigg); \quad \vec{\textbf{\emph{H}}}=\frac{1}{2}\bigg(\vec{\textbf{H}}\mid_{z+}+\vec{\textbf{H}}\mid_{z-}\bigg).
\end{equation}
From eqs. (\ref{Fresnel_micro}) I obtain that the macroscopic electric (magnetic) field is the sum of the incident electric (magnetic) field and the radiation-reaction electric (magnetic) field
\begin{equation}
\label{Macroscopic_fields}
\vec{\textbf{\emph{E}}}(t)=\vec{\textbf{\emph{E}}}_i(t)+\vec{\textbf{\emph{E}}}_R(t); \quad \vec{\textbf{\emph{H}}}(t)=\vec{\textbf{\emph{H}}}_i(t)+\vec{\textbf{\emph{H}}}_R(t).
\end{equation}
Both the local fields and the radiation-reaction fields (eqs. (\ref{E, H}, \ref{P, M}, \ref{LocalE}, \ref{Reaction})) are necessary to deduce the expressions of the electric and magnetic surface susceptibilities
\begin{eqnarray}
\label{susceptibilities}
\chi_e &= &\frac{ \vec{\textbf{\emph{P}}}(t)}{\epsilon_0 \vec{\textbf{\emph{E}}}(t)}=\frac{N \alpha_e \vec{\textbf{\emph{E}}}_{loc}(t)}{\vec{\textbf{\emph{E}}}_i(t)+\vec{\textbf{\emph{E}}}_R(t)}=\frac{N\alpha_e}{1-\frac{C_0\alpha_e}{4\pi a^3}} \nonumber \\ 
\chi_m &= &\frac{ \vec{\textbf{\emph{M}}}(t)}{\vec{\textbf{\emph{H}}}(t)}=\frac{N \alpha_m \vec{\textbf{\emph{H}}}_{loc}(t)}{\vec{\textbf{\emph{H}}}_i(t)+\vec{\textbf{\emph{H}}}_R(t)}=\frac{N\alpha_m}{1-\frac{C_0\alpha_m}{4\pi a^3}}. 
\end{eqnarray} 
The macroscopic electric field and the electric surface susceptibility have exactly the same expressions valid for a 2D crystal \cite{Merano17}.

\subsection{The reflection and transmission coefficients}
I express now the components of the reflected and transmitted waves in term of those of the incident wave, giving
\begin{equation}
\label{Fresnel}
r_s=\frac{E_r}{E_i}=-\frac{ik\chi_e}{ik\chi_e+2}+\frac{ik\chi_m}{ik\chi_m+2}; \quad t_s=\frac{E_t}{E_i}=1-\frac{ik\chi_e}{ik\chi_e+2}-\frac{ik\chi_m}{ik\chi_m+2}.
\end{equation}
 The reflection and the transmission coefficients depend only on macroscopic quantities. The microscopic $\alpha_e$ and $\alpha_m$ enter in these expressions only through $\chi_e$ and $\chi_m$.

\section{Conclusion}
A complete classical description of the optical properties of a metasurface has been presented. Starting from the metasurface structure and the electric and magnetic polarizabilities of the microscopic scatterers, this theory computes both the local fields and the macroscopic fields. It provides also the electric and magnetic surface susceptibilities and the reflected and transmitted fields. 

The approach that has been adopted is analogous to the one developed for the optical response of a dielectric 2D crystal \cite{Merano17} (with null magnetic surface susceptibility) and it is valid also for magnetic 2D crystals \cite{Zhang17, Xiaodong17}. The analogy is possible because an oscillating magnetic dipole produces a magnetic field like the electric field from an oscillating electric dipole and vice-versa.

The radiation-reaction electric and magnetic fields play an essential role in this theory. They derive from the microscopic description and they are proportional respectively to the macroscopic surface polarization and surface magnetization currents. Via the boundary conditions a definition of the macroscopic fields is then possible. Then the radiation-reaction fields complement the local fields to fix $\chi_e$ and $\chi_m$. 

For a dielectric 2D crystal the macroscopic electric field coincides with the transmitted electric field because $\vec{\textbf{\emph{E}}}$ is continuous across the boundary. In a metasurface both the electric and magnetic fields are discontinuous across the boundary. Following \cite{Kuester03} I have defined the macroscopic fields, as the average fields at the metasurfaces. As for a dielectric 2D crystal, the expression of the macroscopic fields is the sum of the incident plus the radiation-reaction fields.

\bibliography{biblio}

\end{document}